\documentclass{article}

    \PassOptionsToPackage{numbers, compress}{natbib}

 \usepackage[creativeai, final]{neurips_2025}

\usepackage[utf8]{inputenc} 
\usepackage[T1]{fontenc}    
\usepackage{hyperref}       
\usepackage{url}            
\usepackage{booktabs}       
\usepackage{amsfonts}       
\usepackage{nicefrac}       
\usepackage{microtype}      
\usepackage{xcolor}         

\usepackage{graphicx}
\usepackage{subfigure}
\usepackage{enumitem}
\usepackage{amsmath}
\usepackage{amssymb}
\usepackage{mathtools}
\usepackage{amsthm}
\usepackage{lipsum}
\usepackage{multirow} 
\usepackage{wrapfig}

\title{Large-Scale Training Data Attribution for \\Music Generative Models via Unlearning}

\author{%
  \textbf{Woosung Choi}$^{1}$\thanks{Equal contributon. Email: \{\textit{first}\_\textit{name}.\textit{last}\_\textit{name}\}@sony.com} \qquad 
  \textbf{Junghyun Koo}$^{1*}$ \qquad 
  \textbf{Kin Wai Cheuk}$^{1*}$ \qquad \\
  \textbf{Joan Serrà}$^{1}$ \qquad 
  \textbf{Marco A. Martínez-Ramírez}$^{1}$ \qquad 
  \textbf{Yukara Ikemiya}$^{1}$ \qquad \\
  \textbf{Naoki Murata}$^{1}$ \qquad 
  \textbf{Yuhta Takida}$^{1}$ \qquad 
  \textbf{Wei-Hsiang Liao}$^{1}$ \qquad 
  \textbf{Yuki Mitsufuji}$^{1,2}$ \qquad \\ \\
  Sony AI$^1$ \quad Sony Group Corporation$^2$
}

\begin{document}

\maketitle

\begin{abstract}
This paper explores the use of unlearning methods for training data attribution (TDA) in music generative models trained on large-scale datasets. TDA aims to identify which specific training data points contributed the most to the generation of a particular output from a specific model. This is crucial in the context of AI-generated music, where proper recognition and credit for original artists are generally overlooked. By enabling white-box attribution, our work supports a fairer system for acknowledging artistic contributions and addresses pressing concerns related to AI ethics and copyright. We apply unlearning-based attribution to a text-to-music diffusion model trained on a large-scale dataset and investigate its feasibility and behavior in this setting. To validate the method, we perform a grid search over different hyperparameter configurations and quantitatively evaluate the consistency of the unlearning approach. We then compare attribution patterns from unlearning with non-counterfactual approaches.
Our findings suggest that unlearning-based approaches can be effectively adapted to music generative models, introducing large-scale TDA to this domain and paving the way for more ethical and accountable AI systems for music creation.
\end{abstract}

\section{Introduction}
Generative AI has demonstrated impressive capabilities across  modalities, including text, images, audio, and video, reshaping artistic creation, as shown by \cite{agarwal2024secure, jiang2024AI, zhang2023loop}.
While democratizing creative work, these advancements have also raised concerns regarding authorship, copyright, attribution, and ethics.
Notably, generative models can unintentionally reproduce copyrighted material, posing risks of intellectual property violations~\cite{carlini2023extracting, somepalli2023understanding}.
As highlighted by \citet{deng2023computational}, these issues are especially pressing in music, where proper attribution and credit to original artists and creators are critical, yet often neglected. To address this, training data attribution (TDA) has emerged as a promising direction to identify which training data points contribute to a model's output, thereby enabling fair crediting.

TDA can be approached in two scenarios based on model access. In the black-box scenario, where the model is inaccessible, corroborative (similarity-based) attribution is typically performed by computing similarity between generated outputs and training data using external feature encoders~\cite{barnett2024exploring, batlle2024towards}. While practical, it relies entirely on each encoder's perspective, which does not necessarily align with the generative model's perspective or its inner workings. In contrast, the white-box scenario assumes access to the model's parameters, enabling attribution methods that directly reflect the model's internal behavior.
An intuitive approach in this setting is based on counterfactual reasoning~\cite{koh2017understanding}, asking how the model's prediction would change if a particular training data point was removed. 
The straightforward solution then to measure the influence of a training data point $\mathbf{x}_i$ is to retrain the whole model without $\mathbf{x}_i$ (leave-one-out retraining). However, this method is computationally unfeasible for large-scale datasets. 
Instead of retraining, \citet{koh2017understanding} and \citet{park2023trak} approximate the change in loss using the influence function~\cite{hampel1974influence}. 
To our knowledge, \citet{deng2023computational} is the first work in music generation that explored influence function-based TDA methods, proposing an algorithmic solution to estimate the impact of individual training data items on the generated music. They applied these methods to a Music Transformer \cite{huang2018music} trained on the MAESTRO dataset \cite{hawthorne2018enabling}, which contains about 200 hours of virtuosic  piano performances.
The same experimental setup was adopted by \cite{deng2024efficient} to further explore ensemble-based TDA approaches.

While the aforementioned methods approximate loss change via the influence function or the ensemble approach, machine unlearning emerges as a promising approach to emulate the counterfactual model. Machine unlearning was proposed by \citet{cao2015towards} to forget a specific data item from a pretrained model. 
Recent studies on TDA have employed gradient ascent to maximize the loss on specific training samples, a process that can be interpreted as unlearning~\cite{ko2024mirrored,wang2024data}. 
To mitigate catastrophic forgetting, \citet{wang2024data} introduced a regularization technique using the Fisher Information Matrix (FIM) when unlearning a sample $\mathbf{x}_i$ from the pretrained model, measuring the change in loss between two different checkpoints as the attribution score.
Although unlearning-based TDA has been actively explored in other domains, its application to music generation remains unstudied.


In this paper, we adapt unlearning-based TDA to measure the attribution score of individual training examples on generated music. We train a latent DiT-based text-to-music generation model~\cite{evans2025stable} on a private dataset comprising 115\,k high-quality music tracks with diverse musical styles, totaling approximately 4,356 hours.
To validate our unlearning-based TDA pipeline, we adopt a self-influence attribution setup to assess whether our unlearning method effectively approximates the counterfactual-based approach.
In this setup, we evaluate the fidelity of our unlearning method to ensure accurate TDA by unlearning a training example and evaluating its influence on other training samples (this could be seen as the case where a model incidentally generated a direct duplicate of a training sample). 
Inspired by~\cite{gandikota2023erasing}, we assess self-influence based on two criteria: (1) the influence of the removed sample must be effectively eliminated, and (2) the model's overall performance must remain stable. 
We detail these two metrics in~\ref{subsec:exp_grid_search}. 
We then leverage these criteria to conduct a grid search and identify optimal configurations for TDA with unlearning.

Using the best configuration  from the self-influence setup, we further analyze test-to-train attribution, comparing the unlearning approach with several other non-counterfactual TDA methods that treat the generative model either as a black-box (similarity-based) or as a white-box (similarity- and gradient-informed), and examine how their influence patterns differ.
To our knowledge, this is the first work that explores TDA on a text-to-music DiT, trained on large dataset of diverse musical styles.

\section{Methodology}

Intuitively, the attribution score can be obtained by measuring the impact of removing a training sample on the model's ability to generate a target output $\hat{\mathbf{z}}$ given the condition $\mathbf{c}_i$. Let $\theta_0$ be the model trained on the full dataset $D = \{\mathbf{z}_i\}_{i=1}^N$, where $\mathbf{z}_i = (\mathbf{x}_i, \mathbf{c}_i)$ is an audio–caption pair, and $\theta_{\setminus\mathbf{z}_i}$ be the model trained without $\mathbf{z}_i$. 
While there are multiple ways to define the attribution scores $\tau\left(\hat{\mathbf{z}}, \mathbf{z}_i\right)$, we adopt the definition based on the changes in loss as in~\cite{wang2024data}, which is defined as

\begin{equation}
\tau\left(\hat{\mathbf{z}}, \mathbf{z}_i\right) = \mathcal{L}(\hat{\mathbf{z}}, \theta_{\setminus\mathbf{z}_i}) - \mathcal{L}(\hat{\mathbf{z}}, \theta_0),
\label{eq:influence}
\end{equation}

However, this leave-one-out method is computationally costly, especially when the dataset is huge and the model training is expensive. A slightly better alternative is to unlearn $\mathbf{z}_i$ from $\theta_0$ to obtain $\theta_{\setminus\mathbf{z}_i}$. Nonetheless, this approach still requires unlearning $\theta_0$ for $N$ times to obtain the exhaustive attribution scores for the whole dataset $D$.

As proposed by \citet{wang2024data}, an even better solution is to approximate Eq.~\ref{eq:influence} with the mirrored influence hypothesis~\cite{ko2024mirrored} by unlearning the generated sample $\hat{\mathbf{z}}$ from a pretrained model $\theta_0$ to obtain an approximation to $\theta_{\setminus\hat{\mathbf{z}}}$
\begin{equation}
\label{eq:influence2}
\tau\left(\hat{\mathbf{z}}, \mathbf{z}_i\right) = \mathcal{L}({\mathbf{z}_i}, \theta_{\setminus\hat{\mathbf{z}}}) - \mathcal{L}({\mathbf{z}_i}, \theta_0).
\end{equation}
This approach requires only a single unlearning step per generated sample, further reducing the computation cost. In the following section, we detail the unlearning algorithm used in this work to obtain $\theta_{\setminus\hat{\mathbf{z}}}$.

\subsection{Unlearning Algorithm}
While the most intuitive way to unlearn \(\hat{\mathbf{z}}\) is to directly maximize its loss, this approach can lead to catastrophic forgetting~\cite{kirkpatrick2017overcoming}. To unlearn the generated sample $\hat{\mathbf{z}}$ without catastrophic forgetting, the unlearned objective should be defined as
\begin{equation}
\label{eq:unlearning_objective1}
\mathcal{L}^{\hat{\mathbf{z}}}_\text{unlearn}(\theta)=-\mathcal{L}(\hat{\mathbf{z}},\theta)+\sum_{\mathbf{z}_i\in D}\mathcal{L}(\mathbf{z}_i,\theta),
\end{equation}
where the first term unlearns $\hat{\mathbf{z}}$ by maximizing the diffusion loss of the generated sample and the second term acts as a regularization to prevent the model from forgetting existing training data by minimizing the diffusion loss for the dataset $D$. Borrowing the idea from~\cite{li2020few}, the second term can be simplified by applying second-order Taylor expansion around $\theta_0$:
\begin{align}\label{eq:unlearning_objective2}
\mathcal{L}(\mathbf{z},\theta) &\approx \mathcal{L}(\mathbf{z},\theta_0) + \nabla_\theta\mathcal{L}(\mathbf{z},\theta_0)^\top(\theta-\theta_0) + \frac{1}{2}(\theta-\theta_0)^\top \mathbf{H}(\theta-\theta_0) \nonumber \\
&\approx \frac{1}{2}(\theta-\theta_0)^\top \mathbf{F}(\theta-\theta_0),
\end{align}
where $\mathcal{L}(\mathbf{z},\theta_0)$ is a constant and the gradient $\nabla_\theta\mathcal{L}(\mathbf{z},\theta_0)$ at $\theta_0$ should be close to zero for a fully trained model, so both terms can be ignored, leaving behind only the last term. It has been proven in existing literature~\cite{barshan2020relatif, grosse2023studying, Lee_2022_CVPR} that the Hessian $\mathbf{H}$ is equivalent to the Fisher information matrix (FIM) $\mathbf{F}$. Plugging Eq.~\ref{eq:unlearning_objective2} back to Eq.~\ref{eq:unlearning_objective1}, we have the following unlearning objective:
\begin{equation}
\mathcal{L}^{\hat{\mathbf{z}}}_\text{unlearn}(\theta)=-\mathcal{L}(\hat{\mathbf{z}},\theta)+\frac{N}{2}(\theta-\theta_0)^\top \mathbf{F}(\theta-\theta_0). 
\label{eq:unlearning_objective}
\end{equation}

Note that $\nabla_\theta\mathcal{L}^{\hat{\mathbf{z}}}_\text{unlearn}(\theta)=0$ when the loss attains its optimal point and $\nabla_\theta(\theta-\theta_0)^\top \mathbf{F}(\theta-\theta_0)=2\mathbf{F}(\theta-\theta_0)$ (the gradient of quadratic form for symmetric $\mathbf{F}$). Taking the gradient of Eq.~\ref{eq:unlearning_objective} w.r.t $\theta$ and rearranging the terms on both sides, we have the following update rule~\cite{wang2024data}:
\begin{align}
0&=-\nabla_\theta\mathcal{L}(\hat{\mathbf{z}},\theta)+N\mathbf{F}(\theta-\theta_0)\nonumber\\
\theta&=\theta_0+\frac{1}{N}\mathbf{F}^{-1}\nabla\mathcal{L}(\hat{\mathbf{z}},\theta).
\label{eq:unlearning_objective_final}
\end{align}
Note that for diffusion models, the loss depends on the denoising timestep $t$. So we calculate the average loss across multiple timesteps $T$, i.e. $\mathcal{L}_(\hat{\mathbf{z}},\theta)=\frac{1}{T}\sum_{t=1}^T\mathcal{L}_t(\hat{\mathbf{z}},\theta)$.


\subsection{Fisher Information Matrix}

The FIM quantifies the amount of information an observation $z$ carries about the model parameters $\theta$, reflecting the curvature of the log-likelihood function.
Mathematically, the FIM is defined as
\begin{equation*}
\mathbf{F} = \mathbb{E}_{\mathbf{z} \sim p(\mathbf{z}|\theta)} \left[ \left( \nabla_\theta \log p(\mathbf{z}\mid \theta) \right) \left( \nabla_\theta \log p(\mathbf{z}\mid \theta) \right)^\top \right].
\end{equation*}
Computing the full FIM is often costly. Thus, a diagonal approximation is commonly used \cite{kirkpatrick2017overcoming}, where each diagonal element $(\mathbf{F}_{\text{diag}})_{jj}$ is estimated by averaging the squared gradients over the $N$ data samples $\mathbf{z}_i$:
\begin{equation*}
(\mathbf{F}_{\text{diag}})_{jj} \approx \frac{1}{N}\sum_{i=1}^N \left( \frac{\partial \log p(\mathbf{z}_i\mid \theta)}{\partial \theta_j} \right)^2.
\end{equation*}
In the context of diffusion models where $\log p(\mathbf{z}_i\mid \theta)=\mathcal{L}_t(\mathbf{z}_i,\theta)$, this is further averaged across $T$ timesteps $t$:
\begin{equation*}
(\mathbf{F}_{\text{diag}})_{jj} \approx \frac{1}{N}\sum_{i=1}^N\frac{1}{T} \sum_{t=1}^{T} \left( \frac{\partial \mathcal{L}_t(\mathbf{z}_i,\theta)}{\partial \theta_j} \right)^2.
\end{equation*}
Now, we have all the information required to unlearn our model via Eq.~\ref{eq:unlearning_objective_final}, and then calculate the attribution score using Eq.~\ref{eq:influence2}.

\subsection{Masking Silence}
\label{subsec:masking_silence}
For music generative modeling, we build upon the DiT model proposed by~\citet{evans2024fast}, which processes variable-length audio as input. Generally, zeros are padded to shorter clips to match the required length. 
Notably, one can choose to apply or omit masking for this padded section. Specifically, a mask $\text{M}$ can be applied to exclude the padded section from loss computation. We assume the model was trained without masking, which is the default setting~\footnote{https://github.com/Stability-AI/stable-audio-tools}. 

Despite this setting, we can apply a mask $\text{M}_\text{U}$ when unlearning a target sample, and a mask $\text{M}_\text{L}$ when computing the loss for measuring attribution.
We also propose a ``mixed'' strategy: applying $\text{M}_\text{U}$ but not $\text{M}_\text{L}$.
$\text{M}_\text{U}$ ensures that the zero-padded section, not part of the actual content, does not interfere with unlearning. 
However, during loss computation, we omit $\text{M}_\text{L}$ to remain  consistent with the training setup. A different setup may result in the model's unpredictable behavior, leading to inaccurate attribution.

\section{Experimental Results}

\subsection{Dataset and Model}
We utilize an in-house dataset consisting of 115\,k high-quality music tracks spanning a diverse range of genres and styles. This dataset is used exclusively to train our base music generative model.
To evaluate unlearning-based attribution, we consider two experimental setups: (1) \textit{Train-to-Train}, which estimates attribution by unlearning individual training samples, and (2) \textit{Test-to-Train}, which attributes generated outputs to specific training data points.
The Train-to-Train setup serves as a controlled way to test the efficacy of the unlearning method by unlearning a training instance and measuring attribution scores to verify whether the unlearned sample is correctly identified as highly influential. In this setup, we select 40 training samples using k-means clustering on CLAP audio embeddings to ensure diversity across the dataset, and use them for the grid search experiments described in Section~\ref{subsec:exp_grid_search}. The Test-to-Train setup is employed in the context of a more qualitative evaluation: we generate 16 two-minute music tracks using distinct text prompts and examine which training data the model attributes to each generated output. This setup is used for comparison with other attribution results, as described in Section~\ref{subsec:test_to_test}.

As mentioned in the previous section, we train a latent diffusion transformer~\citep[DiT;][]{peebles2023scalable} building upon the methodology of Stable Audio~\cite{evans2024fast, evans2025stable}. We first train a variational autoencoder~\citep[VAE;][]{kingmaW2014auto} using the Stable Audio configuration to encode 44.1\,kHz stereo audio into a latent space with a dimensionality of 64 and a time downsampling ratio of 2048.
We employed the $v$-objective diffusion process method \cite{salimans2022progressive} to train our latent DiT. The maximum length of audio that our DiT can process is approximately two minutes, corresponding to 2584 latent frames.
The model is conditioned on CLAP embeddings to enable text-to-music generation, as described in \citet{evans2025stable}. Additionally, it incorporates timing conditions to support variable-length generation, following the methodology outlined in \citet{evans2024fast}.
We computed $\text{FD}_{\text{openl3}}$ on Song Describer reference data \cite{manco2023thesong} to evaluate the overall quality of the generated music, following \citet{evans2025stable}. Our music generative model achieved an $\text{FD}_{\text{openl3}}$ of 110.5, which falls between the performance of Stable Audio 1.0 (142.5) and 2.0 (71.3).

For each unlearning step following Eq.~\ref{eq:unlearning_objective_final}, we average gradients over 2048 random timesteps. A single step takes approximately 20 minutes on an NVIDIA H100 80GB GPU. Computing then the losses for all the training data points requires around 5~hours using 8~H100 GPUs.

\begin{table*}[t]
\centering
  \caption{Grid search results for optimal unlearning hyperparameters. $\text{FD}_\text{openl3}$ is 110.5 for the original checkpoint.}
  \label{Table:grid}
  \resizebox{0.94\linewidth}{!}{%
  \centering
  \resizebox{\textwidth}{!}{%
  \begin{tabular}{ccccccc}
    \toprule
    \textbf{Target Layer} & $\text{M}_\text{U}$ & $\text{M}_\text{L}$ & \textbf{$\text{R}(\text{z}_\text{tar})$} & \textbf{$\text{CLAP}_\text{topk}$} & \textbf{$\text{CLAP}_\text{botk}$} & \textbf{$\text{FD}_\text{openl3}$} \\
    \midrule
    Cross-Attention's \textit{to\_kv} weights & \checkmark &  & 103.2 & 0.38 & 0.35 & 110.5 \\
    Cross-Attention Layers       & \checkmark &    & 1.4  & 0.60 & 0.32 & 110.4 \\
    Self-Attention Layers        & \checkmark &    & 1.1  & 0.63 & 0.30 & 110.5 \\
    All the Transformer Layers   & \checkmark & \checkmark & \textbf{1.0}  & 0.80 & 0.38 & 110.5 \\
    All the Transformer Layers   &  &  & 6615.7  & \textbf{0.82} & 0.42 & 110.5 \\
    All the Transformer Layers   & \checkmark  &  & \textbf{1.0}  & 0.66 & \textbf{0.26} & 110.5\\
    \bottomrule
  \end{tabular}
  }
  }
\end{table*}

\subsection{Self-Influence Experiment and Tuning}
\label{subsec:exp_grid_search}

The employed unlearning method features a number of hyperparameters, such as learning rate, target layers, and number of steps (number of model weights' updates following Eq.~\ref{eq:unlearning_objective_final}). To explore such different options and select the best combination, we performed a grid search. In this grid search, we computed Train-to-Train data attribution, where we unlearn a train data point from the model and compute the attribution score (Eq.~\ref{eq:influence}) for each item in the training set.
Specifically, we examined learning rates ranging from $10^{-7}$ to $10^{-1}$, number of steps from 1 to 4, multiple groups of target layers, and different methods for masking silence (see section~\ref{subsec:masking_silence}). Due to space constraints, we report only the effects of the target layers and the methods for masking silence, as summarized in Table~\ref{Table:grid}, while fixing the learning rate to~$10^{-6}$ and the number of steps to~1 (the best combination we found for these hyperparameters).

In Table~\ref{Table:grid}, we report several metrics to evaluate whether we successfully unlearned the target data while preserving other information. The rank of the target sample, denoted as $\text{R}(\mathbf{z}_{\text{tar}})$, represents the position of the unlearned data point $\mathbf{z}_{\text{tar}}$ in the sorted list of attribution scores $\{\tau(\mathbf{z}_{\text{tar}}, \mathbf{z}_i)\}_{i=1}^n$, where a rank of 1 corresponds to the best attribution score. If $\text{R}(\mathbf{z}_{\text{tar}})$ is greater than 1, it indicates that some samples, which are not the target, were more affected than the target. 
We also report CLAP$_{\text{topk}}$, the mean CLAP cosine similarity of the top-k attribute scores. Specifically, we measure the mean CLAP cosine similarities of  $\mathbf{z}_{\text{tar}}$ and $\mathbf{z}_k$, where $\mathbf{z}_k$ belongs to the top $k$ attribution scores. We hypothesize that unlearning a target track impacts tracks with similar musical components more than irrelevant tracks, making a higher CLAP$_{\text{topk}}$ preferable. We measure the cosine similarities of the top~100 tracks. Similarly, we report CLAP$_{\text{botk}}$, the mean CLAP cosine similarity of 100 tracks with the least significant attribution scores.

As shown in Table~\ref{Table:grid}, omitting both masks did not result in a low $\text{R}(\mathbf{z}_{\text{tar}})$. The $\text{R}(\mathbf{z}_{\text{tar}})$ is particularly high for shorter tracks (typically less than 30\,s) because unlearning these tracks involves unlearning the padded long silence, which may have a greater impact than the actual content.
Enabling both masks prevents this, achieving an $\text{R}(\mathbf{z}_{\text{tar}})$ of 1.0. However, some short tracks appear frequently in different target tracks due to a mismatch between training loss and  attribution loss. Extremely short tracks (less than 10\,s) tend to have high losses for their actual (non-padded) frames because the losses are averaged over the entire frames (120\,s).
To mitigate this problem, we used a mixed strategy, applying only $\text{M}_\text{U}$, which results in the same $\text{R}(\mathbf{z}_{\text{tar}})$ of 1.0.

We also investigated the effect of the target layers. In our experiment, unlearning all the weights in the transformer blocks achieved the highest CLAP$_{\text{topk}}$ and the lowest CLAP$_{\text{botk}}$ among the mixed results. In contrast to \citet{wang2024data}, unlearning only self-attention layers, cross-attention layers, or to\_kv layers in each cross-attention layer was found to be suboptimal (Table~\ref{Table:grid}). 
$\text{FD}_{\text{openl3}}$ did not vary significantly, indicating that the model does not forget information unrelated to the target sample.
We unlearned all the transformer layers with the mixed strategy in the subsequent Test-to-Train attribution experiment.

\begin{figure*}[t]
    \centering
    \includegraphics[width=0.48\textwidth]{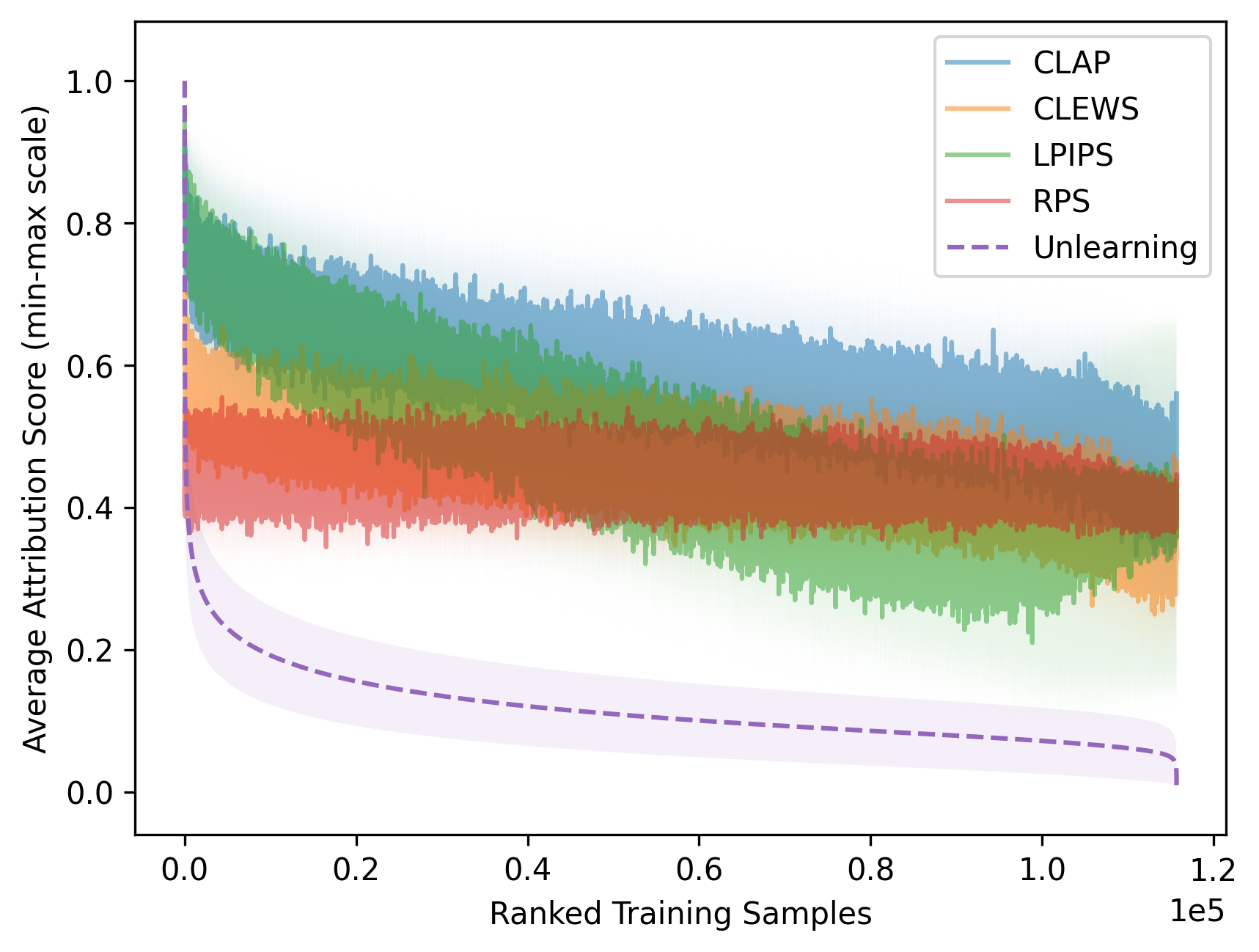}
    \hfill
    \includegraphics[width=0.48\textwidth]{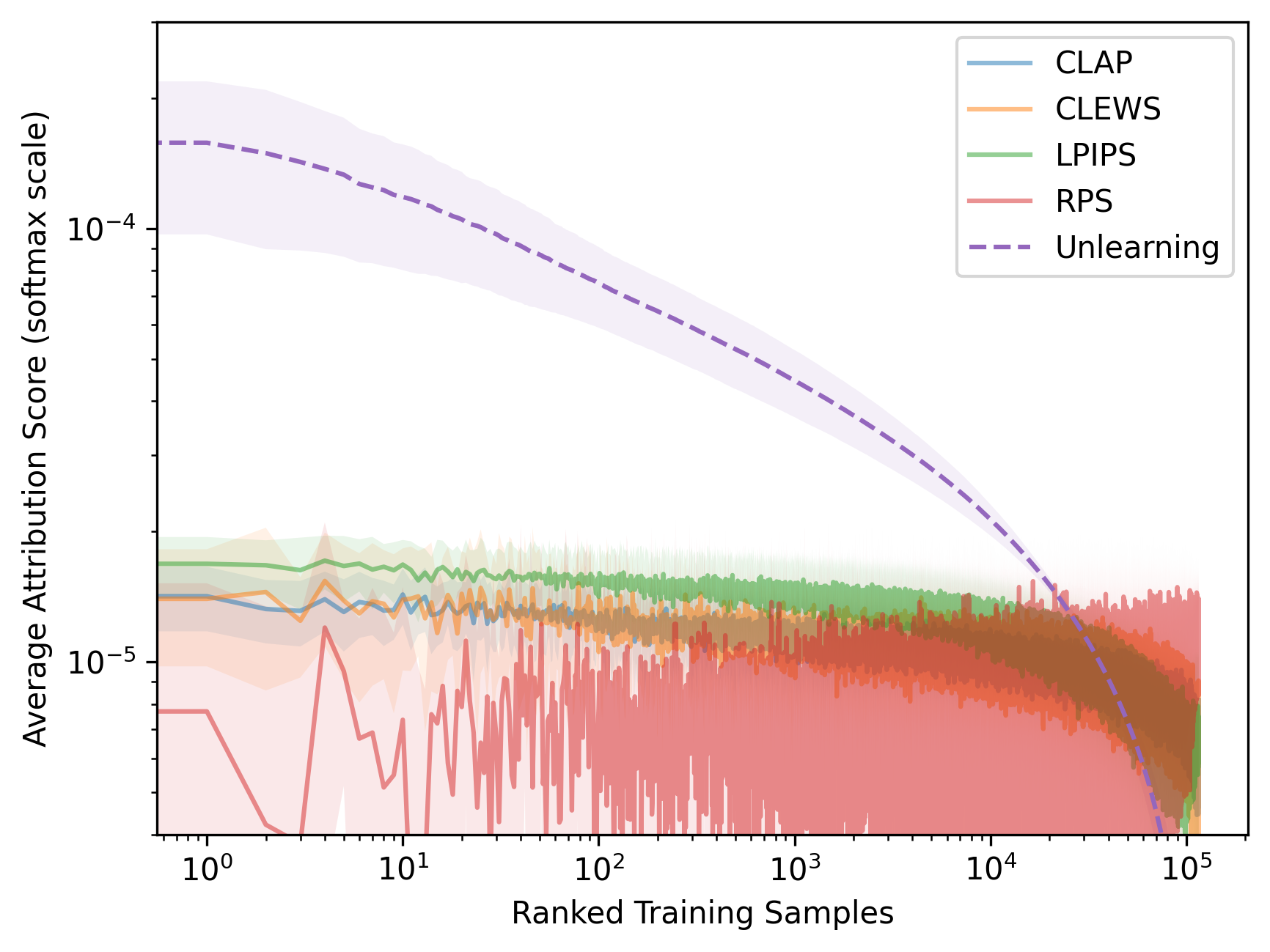}
    \vspace{-0.1cm}
    \caption{Comparison of attribution scores from unlearning- and similarity-based methods. Mean (line) and standard deviation (shading) over attribution scores from 16~generated test samples. Min-max (left) and softmax (right) normalizations are shown (notice the logarithmic axes in the later).}
    \vspace{-0.2cm}
    \label{fig:quota_based_tendency}
\end{figure*}

\subsection{Comparison with Non-counterfactual Methods}
\label{subsec:test_to_test}
We finally compare our unlearning-based attribution method with alternative approaches that do not incorporate counterfactual reasoning: CLAP~\cite{laionclap2023}, CLEWS~\cite{clews}, LPIPS~\cite{lpips}, and RPS~\cite{rps}, under the \textit{Test-to-Train} setup. These methods can all preserve the temporal dimension by windowing the input audio into overlapping segments. Attribution is computed in an \textit{all-against-all} manner, where similarity is computed across all time-wise segments for both target and training tracks and the maximum value is taken as the attribution score.
The \textbf{CLAP} model encodes 10-second audio segments, and we extract embeddings with a hop size of 1~second to preserve temporal resolution across the track.
\textbf{CLEWS}, a contrastive embedding for capturing musical identity across different versions of the same musical piece, follows CLAP in treating the generative model as a black-box.
In contrast, \textbf{LPIPS} utilizes intermediate activations from the generative model by computing similarity at each DiT layer's output and then averaging across all layers.
\textbf{RPS} decomposes the model's pre‑activation prediction $\Phi(\hat{z})$ into a weighted sum over training examples: $\Phi(\hat{z}) = \Sigma_{i=1}^n \tau(\hat{z}, z_i)$. The attribution score of $i^\text{th}$ training sample is $\tau(\hat{z},z_i) = \alpha_i f(z_i)^\top f(\hat{z})$, where $f(\cdot)$ is the feature vector from the last layer of DiT. The representer value $\alpha_i = \frac{1}{-2\lambda n} \frac{\partial \mathcal{L}(z_i, \theta)}{\partial \Phi(z_i, \theta)}$ can be reformulated as $\alpha_i = \frac{1}{-\lambda n} (\Phi(z_i) - z_i)$ as the training objective of our generative model is mean squared error. The final attribution score is computed by taking the average of signed sum of its components.

\begin{wrapfigure}{r}{0.45\textwidth}
    \centering
    \vspace{-0.6cm}
    \includegraphics[width=0.45\textwidth]{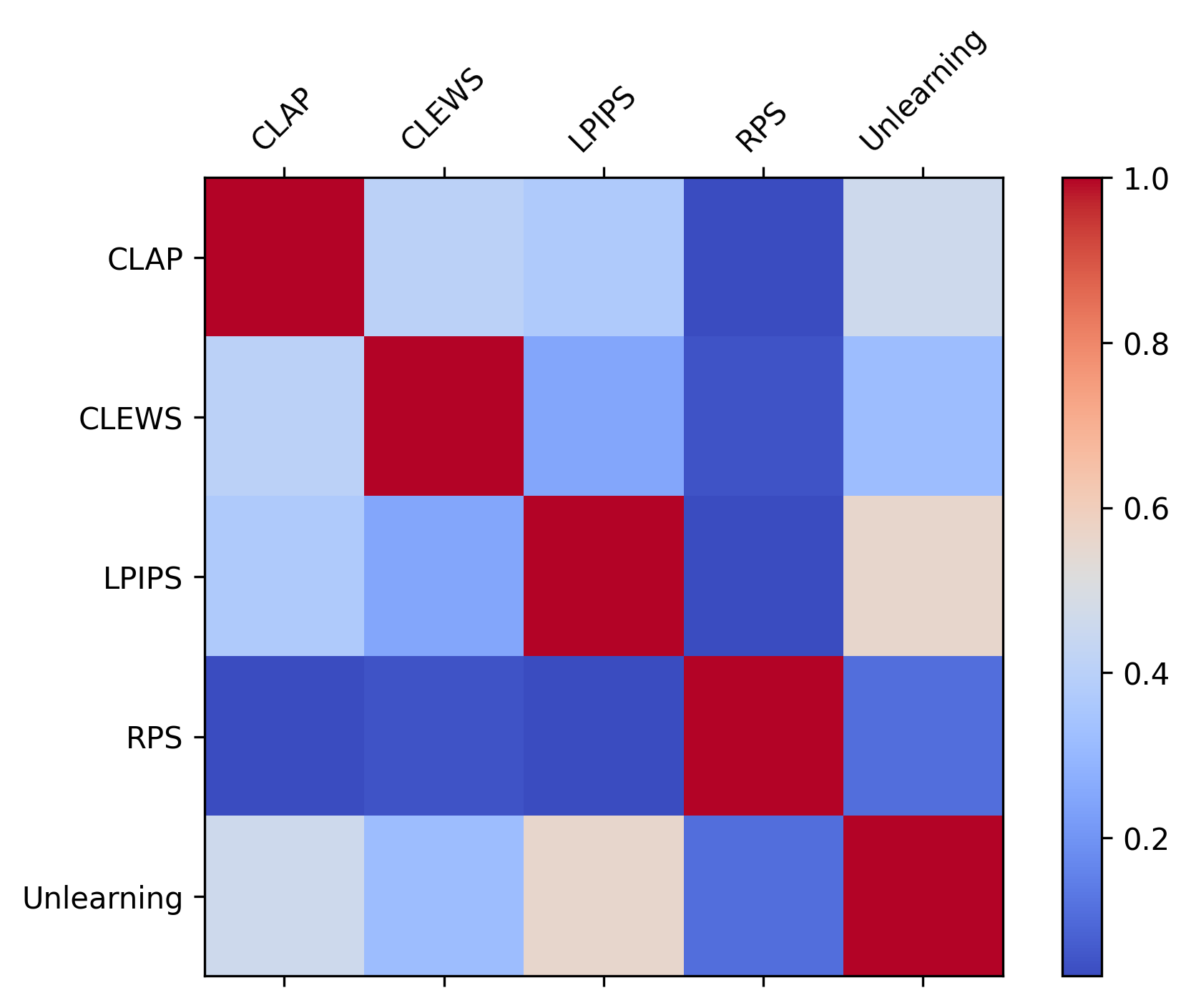}
    \vspace{-0.5cm}
    \caption{Correlation matrix between different attribution methods.}
    \vspace{-0.2cm}
    \label{fig:correlation_matrix}
\end{wrapfigure}

Figure~\ref{fig:quota_based_tendency} visualizes the attribution scores' distribution by sorting all training samples in descending order, based on their unlearning scores, and retrieving the corresponding scores from each of the other methods for the same samples. We present two views: one using min-max normalization to account for scale differences across methods, and the other using softmax scaling (where each method's scores sum to 1) to highlight differences in the overall attribution distribution and sparsity patterns. From these plots, we observe: (1) the unlearning-based method exhibits a sharp concentration of influence in the top few samples, indicating high influence on a small subset of training examples; and (2) the similarity-based methods follow a similar trend but with higher variance and a more gradual decrease in attribution scores, suggesting a broader and less concentrated attribution pattern.
These qualitative patterns are confirmed by the correlation analysis shown in Figure~\ref{fig:correlation_matrix}. Notably, the unlearning-based scores show the strongest Pearson correlation with other attribution methods in the order of LPIPS, CLAP, CLEWS, and RPS, with correlation coefficients of 0.56, 0.46, 0.32, and 0.11, respectively.
This result aligns with methodological similarities: unlearning and LPIPS may exhibit the highest correlation as both leverage internal information from the generative model. Likewise, CLAP and CLEWS also show strong mutual correlation, reflecting their reliance on external embeddings. In contrast, RPS demonstrates low correlation with all other methods, suggesting it captures a distinct attribution pattern.

\section{Conclusion}
This paper presents a practical approach for training data attribution in music generative models using machine unlearning. We apply unlearning techniques to a text-to-music diffusion model trained on a large-scale in-house dataset, and conduct a grid search by unlearning training data itself to identify configurations suitable for attribution. We  compare the unlearning-based results with other attribution methods on generated samples, finding that, while the unlearning and others show similar trends, their attribution patterns differ. This work provides a  framework for applying unlearning-based attribution to music generation models at scale.

\bibliography{reference}
\bibliographystyle{plainnat}

\end{document}